\documentclass{article}
\usepackage{spconf,amsmath,graphicx,amsfonts}
\usepackage{multirow}
\usepackage{color}

\usepackage{listings}
\usepackage{xcolor}
\usepackage{textcomp}

\definecolor{codegreen}{rgb}{0,0.6,0}
\definecolor{codegray}{rgb}{0.5,0.5,0.5}
\definecolor{codepurple}{rgb}{0.58,0,0.82}
\definecolor{backcolour}{rgb}{0.95,0.95,0.92}

\lstdefinestyle{mystyle}{
    backgroundcolor=\color{backcolour},   
    commentstyle=\color{codegreen},
    keywordstyle=\color{magenta},
    numberstyle=\tiny\color{codegray},
    stringstyle=\color{codepurple},
    basicstyle=\ttfamily\footnotesize,
    breakatwhitespace=false,         
    breaklines=true,                 
    captionpos=b,                    
    keepspaces=true,                 
    numbers=left,                    
    numbersep=5pt,                  
    showspaces=false,                
    showstringspaces=false,
    showtabs=false,                  
    tabsize=2
}

\usepackage{hyperref}
\usepackage{enumitem}
\usepackage{booktabs}
\usepackage{multirow}
\usepackage{hhline}
\usepackage{amsmath}
\usepackage[T4,T1]{fontenc}
\makeatletter
\newcommand{\do@openE}[1]{%
  \mbox{\fontsize{#1}\z@\usefont{T4}{cmr}{m}{n}\symbol{130}}%
}
\newcommand{\openE}{\mathord{\mathchoice
  {\do@openE\tf@size}
  {\do@openE\tf@size}
  {\do@openE\sf@size}
  {\do@openE\ssf@size}
}}
\usepackage{ifthen}

\title{Enhanced Sound Event Localization and Detection in Real 360$^\circ$ audio-visual soundscapes}
\name{Adrian S. Roman\sthanks{This work presents preliminary but replicable empirical results. It is published as a non peer-reviewed preprint. Please feel free to contact the corresponding author (romanguz@usc.edu)} \qquad Baladithya Balamurugan \qquad Rithik Pothuganti}
\address{Viterbi School of Engineering, University of Southern California, California, USA}
\begin{document}
%
\maketitle

\section{Abstract}
\label{sec:abtract} 
This technical report details our work towards building an enhanced audio-visual sound event localization and detection (SELD) network. We build on top of the audio-only SELDnet23 model and adapt it to be audio-visual by merging both audio and video information prior to the gated recurrent unit (GRU) of the audio-only network. Our model leverages YOLO and DETIC object detectors. We also build a framework that implements audio-visual data augmentation and audio-visual synthetic data generation. We deliver an audio-visual SELDnet system that outperforms the existing audio-visual SELD baseline.



\section{Introduction}
\label{sec:intro}

A sound event localization and detection (SELD) system generates temporal detection of active sound classes with their corresponding direction of arrival (DoA) around a microphone array \cite{adavanne2018sound}. The spatiotemporal characterization of sound scenes generated by a SELD system has a wide range of applications, such as in audio-visual navigation systems and bio-acoustic monitoring systems \cite{grumiaux2022survey, chen2020soundspaces, cramer2020chirping}.

In real-world environments, a soundscape is a mixture of one or more sound events. A sound event emerges when object interactions generate sound vibrations, e.g., a musician plays a guitar, a person opens a door. Such events virtually never occur as unimodal auditory events but almost always as audio-visual events. SELD depends on the physical characteristics of acoustic scenes to track moving sources, perform sound source distance estimation \cite{kushwaha2023sound, liang2023reconstructing} and account active sources even if they are occluded by another object. With such challenges in mind, it is clear that the visual modality provides useful information to mitigate ambiguities in a SELD task.

Most of the novel research in audio-visual sound source localization aims to learn audio-visual correspondence \cite{mo2022localizing, mo2022closer}. While these models are robust at localizing sound on an image, they are not designed for estimating physical direction of arrival (DoA) in a soundscape. Such models are often trained with audio-visual datasets that do not contain DoA labels and only contain monoaural data\cite{chen2020vggsound}. For this reason, the sound localization performance strongly depends on the video content \cite{wilkins2023bridging}. This makes models prone to erroneous SELD on frames with no audio or uncorrelated audio activity. 

We introduce a visual branch into the audio-only SELDnet23 baseline from the Classification of Acoustic Scenes and Events (DCASE) Challenge 2023 task3. We equip the SELDnet system with state-of-the-art (SOTA) object detectors such as YOLO\footnote{https://github.com/ultralytics/ultralytics} and DETIC \cite{zhou2022detecting}. Additionally, we enable data generation for model training by implementing novel video and audio processing techniques. In summary, our contributions are:

\begin{enumerate}[nolistsep]
\item An implementation of audio-visual data augmentation techniques originally proposed by Wang et al \cite{wang2023four, wang2022nerc}.
\item An audio-visual synthetic data generator for spatial audio and 360$^\circ$ video.
\item A set of audio-visual SELD systems that outperform the existing baselines.
\end{enumerate}

We provide this work as an open source framework available at \url{https://github.com/aromanusc/SoundQ}

\section{Methods}
\label{sec:methods}

\subsection{Datasets} 
\label{sec:datasets}
 We use the Sony-Tau Realistic Spatial Soundscapes 2023 (STARSS23) dataset\footnote{https://zenodo.org/record/7880637} \cite{shimada2023starss23} for development and evaluation. The dataset contains two 4-channel 3-dimensional recording formats: first-order Ambisonics (FOA) and tetrahedral microphone array (tetra), each recorded at a sampling rate of 24kHz and a bit depth of 16 bits. The corresponding video is 360-degree equirectangular with 1920x960 resolution sampled at 29.97 frames-per-second. The data was strongly labeled with spatiotemporal DoA and class labels for a set of 13 target sound classes. Due to the complexities of building a real audio-visual dataset, the development set of STARSS23 is limited (3.8 hours) compared to the large scale data used in the previous iterations of the DCASE Challenge Task3. We tackle the data scarcity by applying three techniques:

\textbf{Data augmentation:} We adopt and implement the audio channel swapping (ACS) and video pixel swapping (VPS) method proposed by Wang et al. \cite{wang2023four, wang2022nerc}. The ACS method increases the amount of DoA representations by performing audio channel rotations within the recordings. In the video domain, the STARSS23 dataset includes 360$^\circ$ video clips with 1920x960 resolution, corresponding to an azimuth range of $[-90^\circ, 90^\circ]$ and an elevation range of $[-180^\circ, 180^\circ]$. The VPS method matches ACS by applying the same transformations at pixel level. For instance, an ACS transformation like $\phi = \phi - \pi/2$,  $\theta = -\theta$, implies rotating the azimuth angle by $-90^\circ$ and inverting the elevation axis. At the pixel level, the VPS transformation slides the azimuth pixels by 1440 pixel points in the negative direction and inverts the elevation axis.

\begin{figure}
\includegraphics[width=8.2cm]{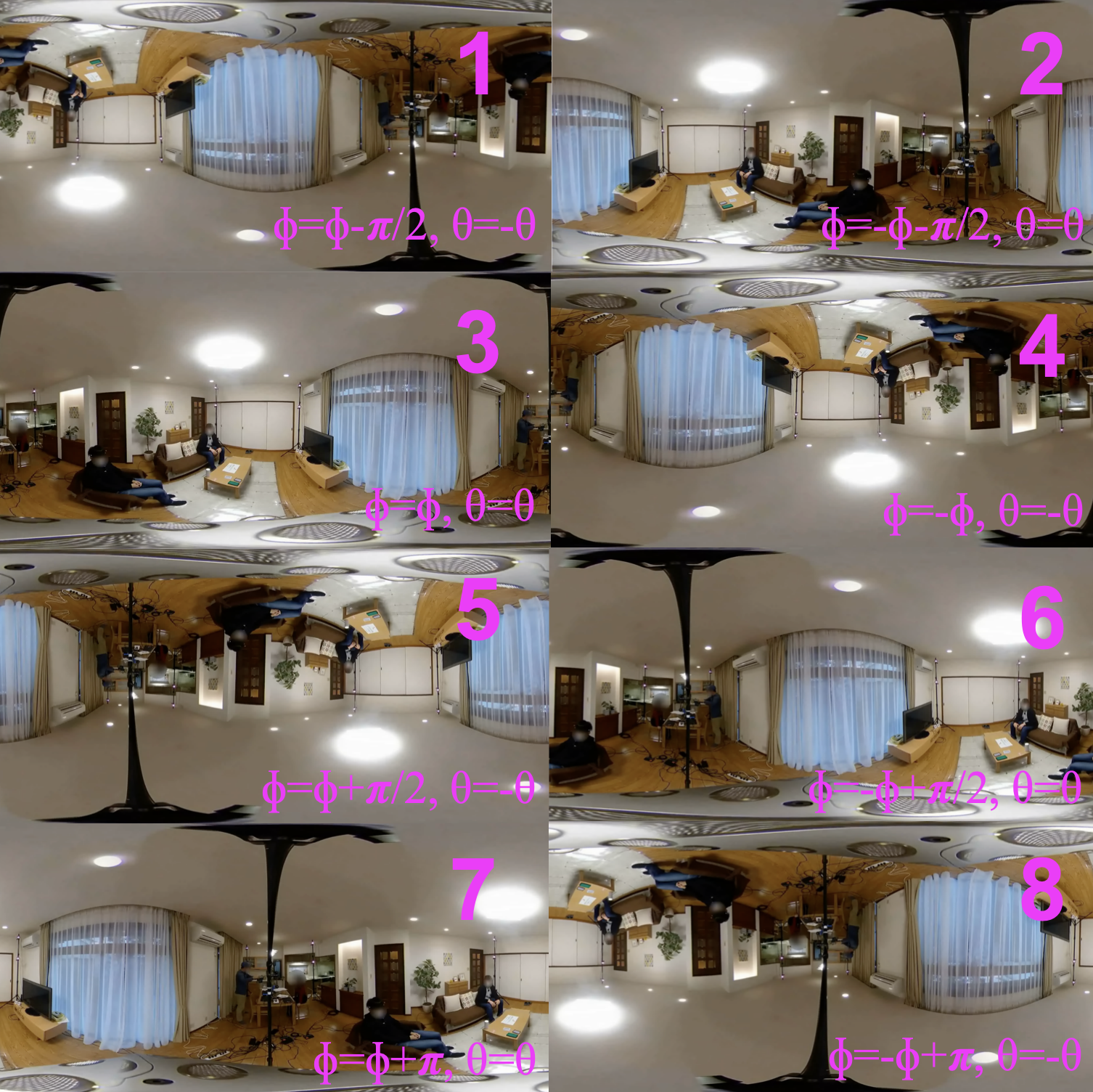}
\caption{Video pixel swapping augmentations.}
\label{fig:augmentation}
\end{figure}

Using the ACS and VPS methods we augment the original audio data by a factor of seven. Figure \ref{fig:augmentation} features the total transformations (including identity) that we integrate in our training set.

\textbf{Audio-only synthetic data:} We leverage the synthetic data from the DCASE Challenge 2022 Task3\footnote{https://zenodo.org/records/6406873}. The recordings were generated through convolution of isolated sound samples with real spatial room impulse responses (SRIRs) captured in nine unique spaces of Tampere University. The training data contains the same sound events classes as the STARSS23 dataset, where the sound assets are sources from the FSD50K dataset \cite{fonseca2021fsd50k}.

\begin{figure}
\includegraphics[width=8cm]{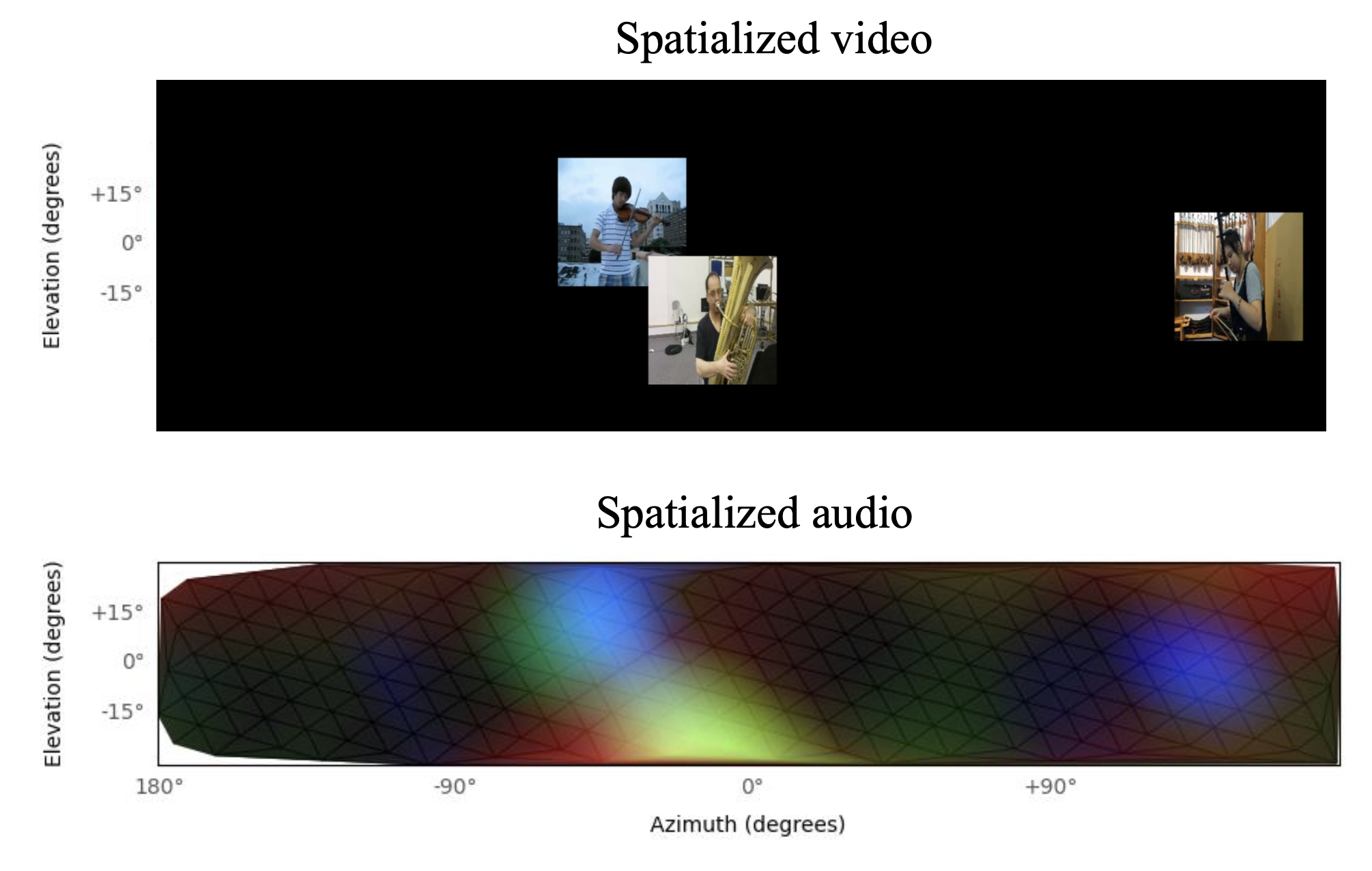}
\caption{$360^\circ$ audio-visual synthetic frame (top) and its spatialized audio displayed using an acoustic camera (bottom).}
\label{fig:synth-data}
\end{figure}

\textbf{Audio-visual synthetic data:} We build a synthetic 360$^\circ$ audio-visual synthetic data generator. We collected a total of 200 YouTube videos containing similar sound events to the STARSS23 dataset. In the audio domain, we generate spatialized sound events using room impulse responses (RIRs) from the METU-SPARG RIR dataset \cite{orhun_olgun_2019_2635758}. The spatial audio synthesizer extracts the audio from a YouTube video and convolves it with an RIR. For the simulated $360^\circ$ video we generate a black video canvas of 1920x960 resolution. We resize the YouTube videos to tiles of 50x50 pixels. We use the 2D projection of the RIR coordinates to superimpose the video frames on the black canvas. The result is a synchronized audio-visual clip. Our audio-visual synthesizer generates variable length audio-visual soundscapes with a maximum of three active events per frame. The metadata format follows the same convetion as the STARSS23 dataset.

We generate a total of 100 synthetic videos each of 30 seconds duration. The top image of Figure \ref{fig:synth-data} shows an example of the generated $360^\circ$ video frames from our synthesizer. We validate the spatialized audio implementation using DeepWave's PyTorch\footnote{https://github.com/adrianSRoman/DeepWaveTorch} to generate acoustic maps \cite{simeoni2019deepwave, roman2024robust}. The color intensity clusters from Figure \ref{fig:synth-data} correspond to the active sound sources, which match the spatial locations of the video overlays. We confirmed the sounds direction of arrival (DOA) estimated a deep acoustic imaging beamformer\footnote{https://github.com/adrianSRoman/DeepWaveDOA}. 


\subsection{The SELD Baseline models} 
\label{sec:baselines}
\textbf{Audio-only baseline:} We use the audio-only SELDnet23 baseline model from the DCASE Challenge Task3 \cite{adavanne2019localization, adavanne2018sound, adavanne2019multi}. The model uses multichannel audio. SELDnet23 is equipped with multi-ACCDOA and it can simultaneously infer presence, class, and spatial coordinates for up to three sound events \cite{shimada2022multi}. Compared to previous SELDnet baselines, it also introduces two multi-head self-attention (MHSA) layers, which makes it more robust for SELD tasks.

\textbf{Audio-visual baseline:} We use the audio-visual SELDnet23 baseline model from the DCASE Challenge Task3 \cite{shimada2023starss23}. This model shared a similar architecture as the audio-only SELDnet, except that it does not contain multi-head self-attention (MHSA) layers. The audio-visual model is equipped with a visual branch that makes use of an object detector (YOLOX \cite{ge2021yolox}). The detector is only used to capture "human" objects. Each bounding box is encoded into two Gaussian-like vectors, corresponding to the azimuth $\rho_{azi}(u_{ij}) \in \mathbb{R}^N$ and elevation $\rho_{ele}(v_{ij}) \in \mathbb{R}^N$. $N$ represents the encoding size, which is $N=37$ for the baseline. Note that the baseline model only supports a maximum of $M=6$ bounding boxes per frame (i.e. $i = 1..6$). The vectors are then concatenated in to a visual embeddings vector $\mathbf{X_v} \in \mathbb{R}^{2 \times M \times N}$. The visual embeddings $\mathbf{X_v}$ are then concatenated with audio embeddings $\mathbf{X_a}$ and sent through an audio-visual decoder network that allows to output multi-ACCDOA labels. 

\begin{figure}
\includegraphics[width=8.5cm]{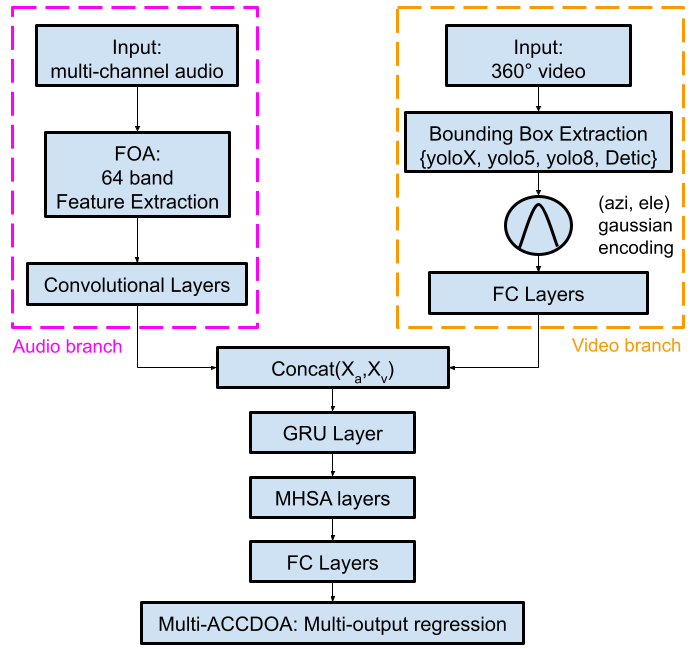}
\caption{Enhanced audio-visual SELD system.}
\label{fig:av-seld}
\end{figure}

\subsection{Audio-visual SELD model enhancements}    
\label{sec:enhance}

The audiovisual SELDnet baseline shows limited performance compared to the audio-only baseline \cite{shimada2023starss23}. Such shortcomings are due to the scarce training data and the architectural limitations of its visual branch \cite{shimada2023starss23}. 

We focus our study towards architectural enhancements to the vision branch and deliver two model variants that replace the original object detector: YOLO-based and DETIC-based. This two-fold approach allows us to understand how enhancing object detection contributes to better SELD performance. Similarly, we are able to understand the contributions of fixed vocabulary detectors (i.e YOLO), compared to detectors with large and customizable vocabulary (i.e DETIC). The YOLO-based detectors attain detection fixed to the COCO dataset classes \cite{lin2014microsoft}. DETIC has the ability to identify 21,000 object classes with strong accuracy as it was trained with a variety of datasets, among those the LVIS dataset \cite{gupta2019lvis}. Furthermore, since DETIC employs a CLIP embedding vector, it is possible to change the model's vocabulary to a customized one. In our case, we customized DETIC's vocabulary to target the STARSS23 classes. We also equip our audio-visual SELD system with two MHSA layers inspired from the audio-only baseline \cite{sudarsanam2021assessment}. Our architectural enhacements therefore integrate either YOLO5, YOLO8 or DETIC as the object detectors. The visual embeddings follow the same format as the audio-visual baseline described in section \ref{sec:baselines}. Figure \ref{fig:av-seld} displays our proposed audio-visual SELDnet23 architecture.

\subsection{Metrics}    
\label{sec:metric}

We employ the SELD metrics proposed by the DCASE Challenge \cite{politis2020overview}. Two metrics relate to DoA estimation: F1-score (F$_{20^\circ}$) and error rate (ER$_{20^\circ}$). F$_{20^\circ}$ is calculated from location-aware precision and recall. ER$_{20^\circ}$ is the sum of insertion, deletion and substitution errors divided by the total number of inferred audio frames. The remaining two metrics relate to class-aware localization: localization error (LE) in degrees and localization recall (LR). LE is the average angular difference between each class prediction and its ground truth labels. LR is the true positive rate of instantaneous detections out of the total annotated sounds.

\subsection{Training procedure}    
\label{sec:training}

Our proposed SELD systems adopt the same hyper-parameters as the audio-only SELDnet23 baseline. We only augment the STARSS23 train split and our synthetic audio-visual data using Wang et al's data augmentation approach \cite{wang2023four, wang2022nerc}. We also include the audio-only synthetic data from DCASE22 discussed in section \ref{sec:datasets}. Audio-visual models were trained  with STARSS23, synthetic audio-visual data, and audio-only synthetic data. For the audio-only synthetic data a black video stream is presented as background. The audio-only model was trained with the augmented STARSS23 and audio-only synthetic data. All models were trained over 100 epochs and the best of 5 validation performances was chosen. 

\begin{figure}
\includegraphics[width=8.7cm]{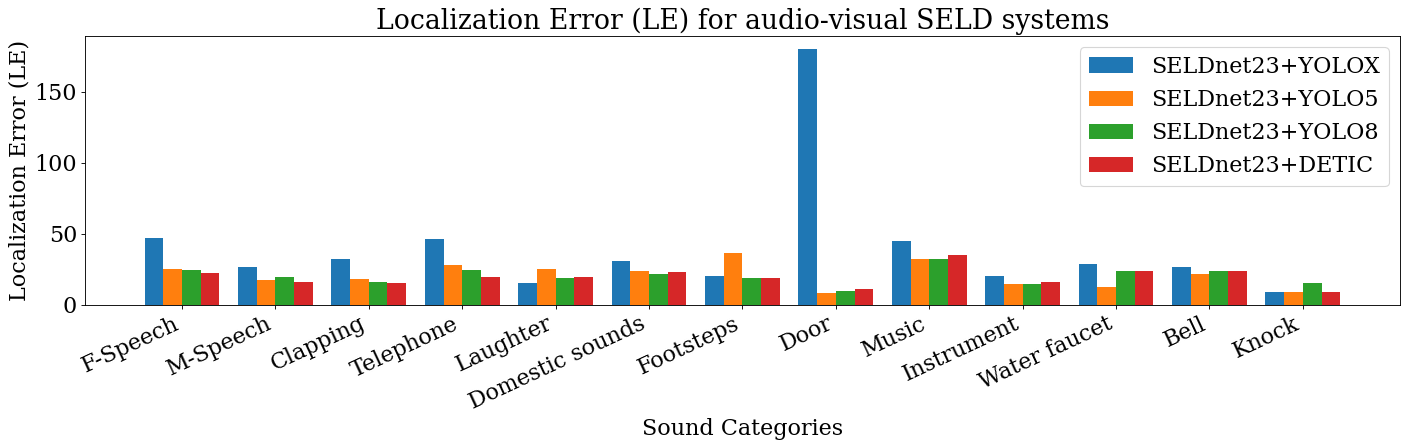}
\caption{Per-class localization error performance comparing the audio-visual baseline against our proposed enhancements.}
\label{fig:benchmark}
\end{figure}

\begin{table*}
 \begin{center}
 \begin{tabular}{llllcccc}
   \toprule\toprule
    Model & detector & data aug & input & ER$_{20^\circ}$ & F$_{20^\circ}$ & LE$^\circ$ $\downarrow$ & LR $\uparrow$ \\ 
   \midrule\midrule
    Baseline AO SELDnet23 & N/A & Yes & FOA + Multi-ACCDOA & 0.57 & 29.9 &  21.6 & \textbf{47.7} \\ \hline
    Baseline AV SELDnet23 & YOLOX & No & FOA + Video & 1.07 & 14.3 & 48.0 & 35.5 \\ \hline
    Baseline AV SELDnet23 & YOLOX & Yes & FOA + Video & 1.37 & 15.0 & 40.62 & 40.0 \\ \midrule\midrule
    \textbf{Ours AV SELDnet} & YOLO5 & Yes & FOA + Multi-ACCDOA + Video & 0.64 & 27.5 & 21.0 & 41.4 \\ \hline
    \textbf{Ours AV SELDnet} & YOLO8 & Yes & FOA + Multi-ACCDOA + Video & 0.63 & 30.9 & 20.3 & 46.1 \\ \hline
    \textbf{Ours AV SELDnet} & DETIC & Yes & FOA + Multi-ACCDOA + Video & 0.64 & 30.6 & \textbf{19.5} & 43.7 \\ \hline
   \bottomrule
   \end{tabular}
\end{center}
\setlength{\belowcaptionskip}{-15pt}
 \caption{Comparison of the baseline SELDnet23 audio-only  and audio-visual against our proposed audio-visual SELDnet architecture. Best performance in the "test-split" from the development STARSS23 dataset. AV=audio-visual, AO=audio-only.}
 \label{tab:starss_results}
\end{table*}

\section{Results} 
\label{sec:results}

Table \ref{tab:starss_results} shows performance by our proposed audio-visual SELDnet enhancements and compares them against the audio-only and audio-visual SELDnet23 baselines. The results below are the evaluation metric scores using the `test split' of the STARSS23 development set. Most models were trained using audio-visual data augmentations (see section \ref{sec:training}), otherwise denoted by `No' on the third column. 

The top three rows correspond to the audio-only (AO) and the audio-visual (AV) SELDnet23 baselines. The AO SELDnet23 features more robust LE and LR metrics compared to the AV SELDnet23. This is because the AV SELDnet23 was only trained on the STARSS23 development set with no data augmentations, while the AO SELDnet23 was trained with a combination of the STARSS23 development set recordings and synthetic audio recordings from DCASE22. Additionally, the AO system is equipped with two MHSA layers that make the system more robust at localization tasks. The third row shows the benefits of data augmentations on the AV SELDnet23, giving a clear improvement on LE \textasciitilde$8^\circ$ and LR \textasciitilde$5^\circ$ over the original AV SELDnet23.

The models equipped with YOLO5 and YOLO8 detectors outperform in all metrics the original AV baseline that uses the tiny YOLOX version. This shows the advantage of using more robust object detectors that are resilient to the equiangular projection distortion on 360$^\circ$ video. The model equipped with YOLO8 outperforms other YOLO-based architectures. This is a product of YOLO8's transformer-based architecture, which has a much higher accuracy compared to earlier YOLO models. Note that the YOLO-based systems inherit object detection from the COCO dataset classes \cite{lin2014microsoft}; hence, because the COCO classes do not overlap with the STARSS23 classes, the systems virtually detects only the "person" class.

The DETIC-based system (last row) is the best-performing model in regards to LE. This model shows the advantage of having a detector with a vocabulary tailored to the STARSS23 classes. The slight degradation in LR performance is an indicator of the added complexity for handling multiple object class detection in the video stream. 

Figure \ref{fig:benchmark} breaks down the LE performance for each sound category, comparing the baseline system (SELDnet23+YOLOX) against our proposed architectural enhancements. In this plot, all models were trained with the augmented training set. It is clear that our proposed model enhancements outperform the audio-visual baseline nearly for all sound categories. In addition, the DETIC-based system generally performs better on classes where there may be clear object-human interactions.

\section{Conclusion and future work} 
\label{sec:conclusion}

We have proposed a set of audio-visual models that outperform the existing audio-only and audio-visual SELDnet23 baselines. Our experiments reveal the benefits of using audio-visual data augmentations and audio-visual synthetic data. We also shed light on the advantages of using SOTA object detectors for fixed vocabulary (e.g., YOLO) and customized vocabulary (e.g., DETIC). While our validation results show enhanced performance over baselines, there may still be over-fitting or domain-shift risks. 

Future work could build on our proposed audio-visual synthetic data generator. There could be large benefits from collecting $360^\circ$ video samples to generate synthetic high-quality audio-visual soundscapes at scale. Improvements to our method could include superimposing video frames using 3D coordinates (including depth), which could used for training SELD systems for depth estimation. Our proposed audio-visual synthetic generation is planned to be integrated in a future release of the Spatial Scaper library \cite{roman2024spatial}.

\section{Acknowledgements} 
This work was done as a final project for CSCI677. The authors thank Prof. Yue Wang for his support and CARC at USC for providing HPC resources for this investigation.

\clearpage 
\vfill\pagebreak

\bibliographystyle{IEEEbib}
\bibliography{refs, strings}
\end{document}